\documentclass[twocolumn,preprintnumbers,amsmath,superscriptaddress,amssymb,aps]{revtex4-1}

\usepackage{graphicx}
\usepackage{dcolumn}
\usepackage{bm}
\usepackage{amsmath}
\usepackage{amssymb}
\usepackage{graphicx}
\usepackage{indentfirst}
\usepackage{booktabs}
\usepackage{multirow}
\usepackage{colortbl}

\selectfont

\begin{document}
\title{Multifield Induced Antiferromagnet Transformation into Altermagnet and Realized Anomalous Valley Hall Effect in Two-dimensional Materials}
\author{Hanbo Sun}
\address{State Key Laboratory for Mechanical Behavior of Materials, School of Materials Science and Engineering, Xi'an Jiaotong University, Xi'an, Shaanxi, 710049, People's Republic of China}
\author{Pengqiang Dong}
\address{State Key Laboratory for Mechanical Behavior of Materials, School of Materials Science and Engineering, Xi'an Jiaotong University, Xi'an, Shaanxi, 710049, People's Republic of China}
\author{Chao Wu}
\address{State Key Laboratory for Mechanical Behavior of Materials, School of Materials Science and Engineering, Xi'an Jiaotong University, Xi'an, Shaanxi, 710049, People's Republic of China}
\author{Ping Li}
\email{pli@xjtu.edu.cn}
\address{State Key Laboratory for Mechanical Behavior of Materials, School of Materials Science and Engineering, Xi'an Jiaotong University, Xi'an, Shaanxi, 710049, People's Republic of China}
\address{State Key Laboratory of Silicon and Advanced Semiconductor Materials, Zhejiang University, Hangzhou, 310027, People's Republic of China}
\address{State Key Laboratory for Surface Physics and Department of Physics, Fudan University, Shanghai, 200433, People's Republic of China}

\date{\today}

\begin{abstract}
Altermagnetism, as a new category of collinear magnetism distinct from traditional ferromagnetism and antiferromagnetism, exhibits the spin splitting without net magnetization. Currently, researchers are focus on searching three-dimensional altermagnetism and exploring its novel physical properties. However, there is a lack of understanding of the physical origin of two-dimensional altermagnetic emergent behavior. Here, we propose an approach to realize the transition from Neel antiferromagnetism to altermagnetism in two-dimensional system using an electric field, Janus structure, and ferroelectric substrate. In monolayer VPSe$_3$, we demonstrate that multiple-physical-fields cause the upper and lower Se atoms unequal to break $\emph{PT}$ symmetry, resulting in altermagnetic spin splitting. Noted that monolayer VPSe$_3$ produces a spontaneous valley splitting of 2.91 meV at the conduction band minimum. The electric field can effectively tune the valley splitting magnitude, while the Janus structure not only changes the valley splitting magnitude, but also alters the direction. More interestingly, when the ferroelectric polarization of Al$_2$S$_3$ is P$\uparrow$, the direction of valley polarization is switched and the magnitude is almost unchanged. However, the valley splitting significantly increases under the P$\downarrow$. It is worth noting that the ferroelectric polarization can switch altermagnetic effect and realize anomalous valley Hall effect. Besides, we reveal the microscopic mechanism of valley splitting by an effective Hamiltonian. Our findings not only provide a method to designing altermagnet, but also enriches the valley physics.
\end{abstract}

\maketitle
\section{Introduction}
Recently, a new type of magnetism named "altermagnetism" has been proposed in condensed matter physics \cite{1,2,3,4,5}. The altermagnetism has two fascinating properties, which mainly manifested as a zero net magnetic moment and spin splitting along specific high symmetry paths without the spin-orbit coupling (SOC). The discovery has not only promoted the development of spintronics, but also enriched the application scenarios of magnetic materials. For example, giant and tunneling magnetoresistance effect is raised in altermagnets \cite{6}. Besides, the anomalous Hall effect is reported in the altermagnet RuO$_2$, which the magnitude can be compared to that of ferromagnet (FM) \cite{7}. In addition, a new type of torque, spin splitter torque, has been theoretically proposed \cite{8} and experimentally observed \cite{9,10} in the altermagnet. Moreover, the heterojunctions of altermagnet are reported to possess chiral Majorana Fermion or Majorana zero energy modes \cite{11,12}. However, these investigations have mainly focused on bulk materials \cite{2,3,4,5,6,7,8,9,10,13,14,15,16}, while two-dimensional (2D) altermagnets pay limited attention. How to tune the 2D antiferromagnet (AFM) into altermagnet is even less involved. But it is crucial to understanding the origin of the emergent behavior.

In spintronics, a new degree of freedom, valley, has been proposed as the third degree of freedom beyond the electron's charge and spin \cite{17,18,19,20}. The valley indicates a local energy minimum or maximum point in the conduction or valence band. At present, the investigation focus of valleytronics is how to realize spontaneous valley polarization and effectively tune \cite{21,22,23,24,25,26,27,28,29}. There are two main ways to achieve spontaneous valley polarization, which is named ferrovalley material \cite{21}. One approach is to break the time-reversal symmetry ($\emph{T}$) with the FM or AFM \cite{21,22,23,24,25,29}, while the other way is to break the inversion symmetry ($\emph{P}$) by ferroelectricity \cite{27,28,30}. It is well known that altermagnets break the combined symmetry of the $\emph{P}$ and $\emph{T}$ (named the $\emph{PT}$ symmetry). Whether the altermagnetism can realize spontaneous valley polarization, and how it differs from the FM and AFM systems?

In this work, we propose a new scheme for achieving altermagnetism in 2D system. Here, we focus on the transformation of AFM into altermagnet by the multifield induction, such as an electric field, Janus structure, and ferroelectric substrate. It is well known that the AFM hold the $\emph{PT}$ symmetry. The multiple-physical-fields can induce the $\emph{PT}$ symmetry breaking, which causes the band degeneracies of spin up and spin down bands to disappear on special high symmetrical path, exhibiting altermagnetic characteristics. Based on the density functional theory (DFT) calculations, we demonstrate this mechanism and phenomena in monolayer VPSe$_3$. Moreover, monolayer VPSe$_3$ exhibits spontaneously valley polarization due to the combined effects of SOC and $\emph{T}$ symmetry breaking. The electric field can effectively tune the magnitude of valley splitting, while the Janus structure (built-in electric field) and ferroelectric substrate can regulate not only the magnitude but also the direction and position of valley polarization. In addition, the ferroelectric polarization can switch altermagnetic effect and realize anomalous valley Hall effect. Our work provide a new direction for investigating altermagnetism and valleytronic devices.

\section{STRUCTURES AND COMPUTATIONAL METHODS}
Based on the framework of the DFT, we employed the Vienna $Ab$ $initio$ simulation package (VASP) \cite{31,32} to investigate the electronic and magnetic properties. The generalized gradient approximation (GGA) with the Perdew-Burke-Ernzerhof (PBE) functional is used to describe the exchange correlation energy \cite{33}. The kinetic energy cutoff for plane-wave basis is set to be 500 eV. A vacuum of 30 $\rm \AA$ is added along the $\emph{c}$-axis, to avoid the interaction between the sheet and its periodic images. The convergence criteria of the total energy and the force for lattice optimization are set to 10$^{-6}$ eV and -0.005 eV/$\rm \AA$, respectively. To describe strongly correlated 3$\emph{d}$ electrons of V, the GGA+U approach is performed with the effective U value (U$_{eff}$ = U - J) of 3 eV \cite{34}. The zero damping DFT-D3 approach of Grimme is considered for the van der Waals (vdW) correction in VPSe$_3$/Al$_2$S$_3$ heterostructure \cite{35}. In addition, to study the Berry curvature, the maximally localized Wannier functions (MLWFs) are used to construct an effective tight-binding Hamiltonian by Wannier90 code \cite{36,37}.

\section{RESULTS AND DISCUSSION }	
\section{RESULTS AND DISCUSSION }	
\subsection{Structure and symmetry}
As shown in Fig. 1(a), it shows the crystal structure of monolayer VPSe$_3$, which is six Se atom nearest to each V atom. The monolayer VPSe$_3$ shows a hexagonal honeycomb lattice with the point group of D$_{3d}$ and space group of P$\bar{3}$1m. The optimized lattice constant of VPSe$_3$ is 6.24 $\rm \AA$. When magnetic order isn't considered, monolayer VPSe$_3$ has the $\emph{P}$ symmetry. However, since the magnetic ground state of VPSe$_3$ is N$\acute{e}$el AFM, both the $\emph{P}$ and $\emph{T}$ symmetry are broken. Despite this, it exhibits invariance when the spatial inversion and time reversal occur simultaneously, that is, the $\emph{PT}$ symmetry (see Fig. 1(b)). It is well established that a system with $\emph{PT}$ symmetry exhibits Kramers degeneracy in its band structure. Namely, as shown in Fig. 1(c), the spin up and spin down bands are degenerate. Besides, the opposite spin sublattices are jointed by the combined rotation and mirror symmetry, denoted as [C$_2$$\|$M]. The [C$_2$$\|$M] symmetry lead to E(s, $\textbf{k}$) = [C$_2$$\|$M]E(s, $\textbf{k}$) = E(-s, M$\textbf{k}$), which plays a vital role in realizing altermagnetism. Here, the s, $\textbf{k}$, and E(s, $\textbf{k}$) represent the spin, momentum, and spin- and momentum-dependent bands, respectively.

\begin{figure}[htb]
\begin{center}
\includegraphics[angle=0,width=1.0\linewidth]{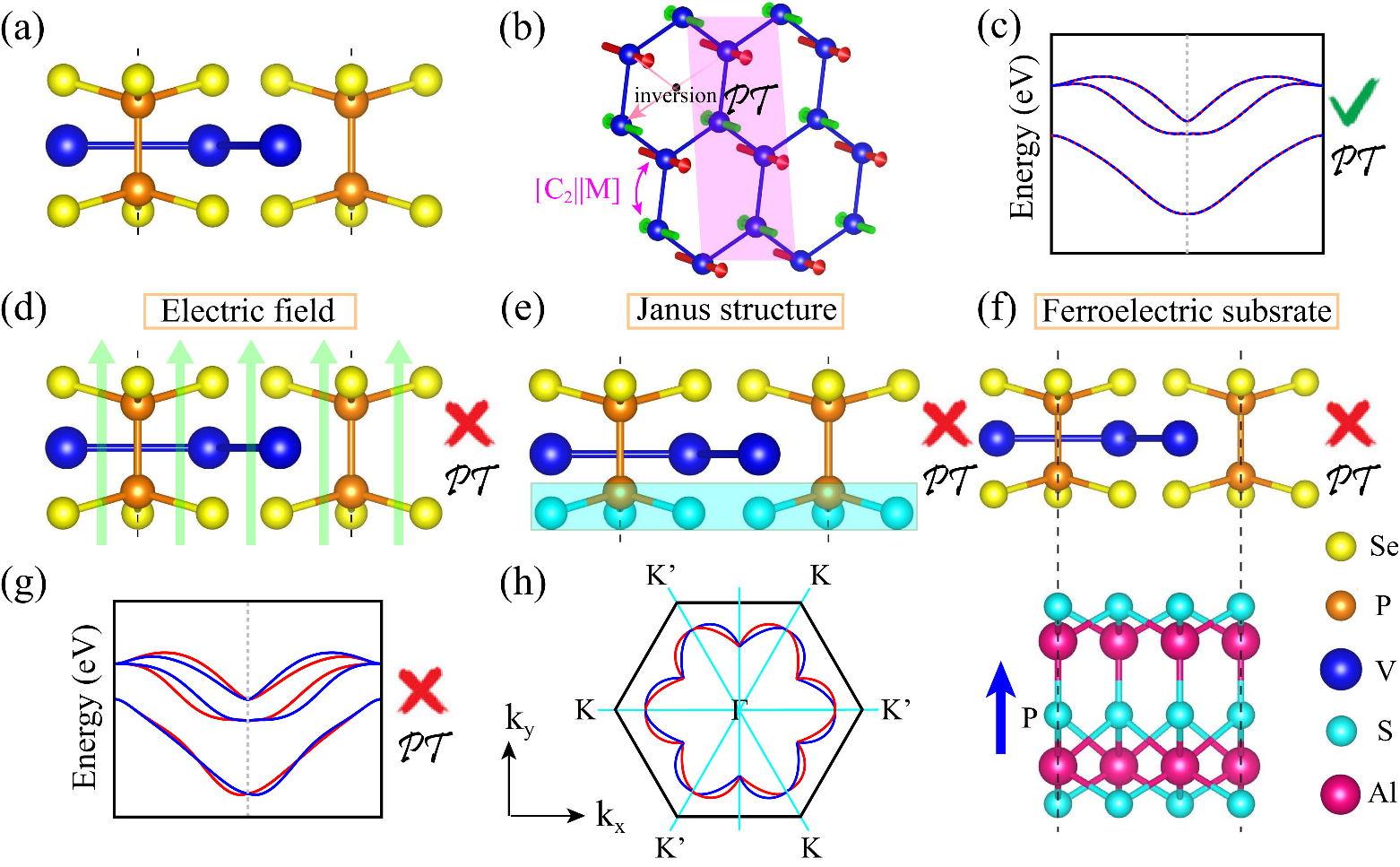}
\caption{(a) The side view of the crystal structure for monolayer VPSe$_3$. (b) Schematic diagram of N$\acute{e}$el type magnetic order for monolayer VPSe$_3$ with the $\emph{PT}$ symmetry. (c) Band structure of AFM-N state with the PT symmetry. Three kinds of field induced $\emph{PT}$ symmetry breaking. (d) Applied an external electric field. (e) Functionalizing the crystal structure into Janus configuration form by substituting the bottom Se layer through S atoms. (f) Ferroelectric substrate is introduced to form vdW heterostructure. (g) Band structure of AFM-N state with broken the $\emph{PT}$  symmetry. (h) The band structures shows i-wave symmetry without the SOC effect. The magenta, the light blue, blue, orange, and yellow balls represent Al, S, V, P, and Se elements, respectively.
}
\end{center}
\end{figure}

Here, we propose three kinds of field to break the $\emph{PT}$ symmetry. As shown in Fig. 1(d-f), these include the application of out-of-plane external electric fields, the functionalizing crystal structure into Janus configuration form by substituting the bottom Se layer through S atoms, and the applied ferroelectric substrate. A common feature of these three methods is that the mirror symmetry can be preserved. Consequently, as shown in Fig. 1(g), the Kramers degeneracy is lifted, and the band structure characteristics of altermagnet are exhibited. More interestingly, as shown in Fig. 1(h), the i-wave altermagnets are formed.

\subsection{Band structure and valley splitting}
To determine the magnetic ground state of monolayer VPSe$_3$, the two most likely magnetic configurations including the FM and Neel type AFM are considered (the Neel Type AFM is hereinafter referred to as AFM). We calculate that the FM state is 277.67 meV more energy than the AFM state, which indicates that the AFM phase is the magnetic ground state of monolayer VPSe$_3$. For the 2D magnetic materials, the out-of-plane magnetic anisotropy energy (MAE) is the basis for their stable existence \cite{38,39}. The MAE is defined as MAE = E$_{100}$ - E$_{001}$, where E$_{100}$ and E$_{001}$ represent the total energy of the magnetic moment along [100] and [001] axis, respectively. The calculated MAE is 0.10 meV, which indicates the easy magnetization direction along the [001] axis.

Then, we calculate the spin-polarized band structure of monolayer VPSe$_3$. As shown in Fig. 2(a), the spin up and spin down bands are degenerate, which the $\emph{PT}$ symmetry induces the Kramers degeneracy. It is worth noting that the K/K' valley of conduction band minimum (CBM) is degenerate. Noted that Fig. 2(c) plots the valence band at around -1.2 eV. When the SOC is switched on, as shown in Fig. 2(b), the energy valley degeneracy of K and K' disappears, resulting in 2.91 meV spontaneous valley splitting. Here, we define valley splitting as the energy difference between K and K' points. The origin of valley splitting in monolayer VPSe$_3$ is consistent with the previously reported magnetic system due to the combined $\emph{T}$ symmetry broken and SOC effect. To understand the orbital composition of valley splitting, we calculate the orbital-resolved band structure of monolayer VPSe$_3$. As shown in Fig. S1, the valence band maximum (VBM) bands are mainly contributed by the d$_{x^2-y^2}$+d$_{xy}$+d$_{z^2}$ orbitals of V atoms, while the CBM bands are dominated by the p$_{x}$+p$_{y}$ orbitals of Se atoms.

\begin{figure}[htb]
\begin{center}
\includegraphics[angle=0,width=1.0\linewidth]{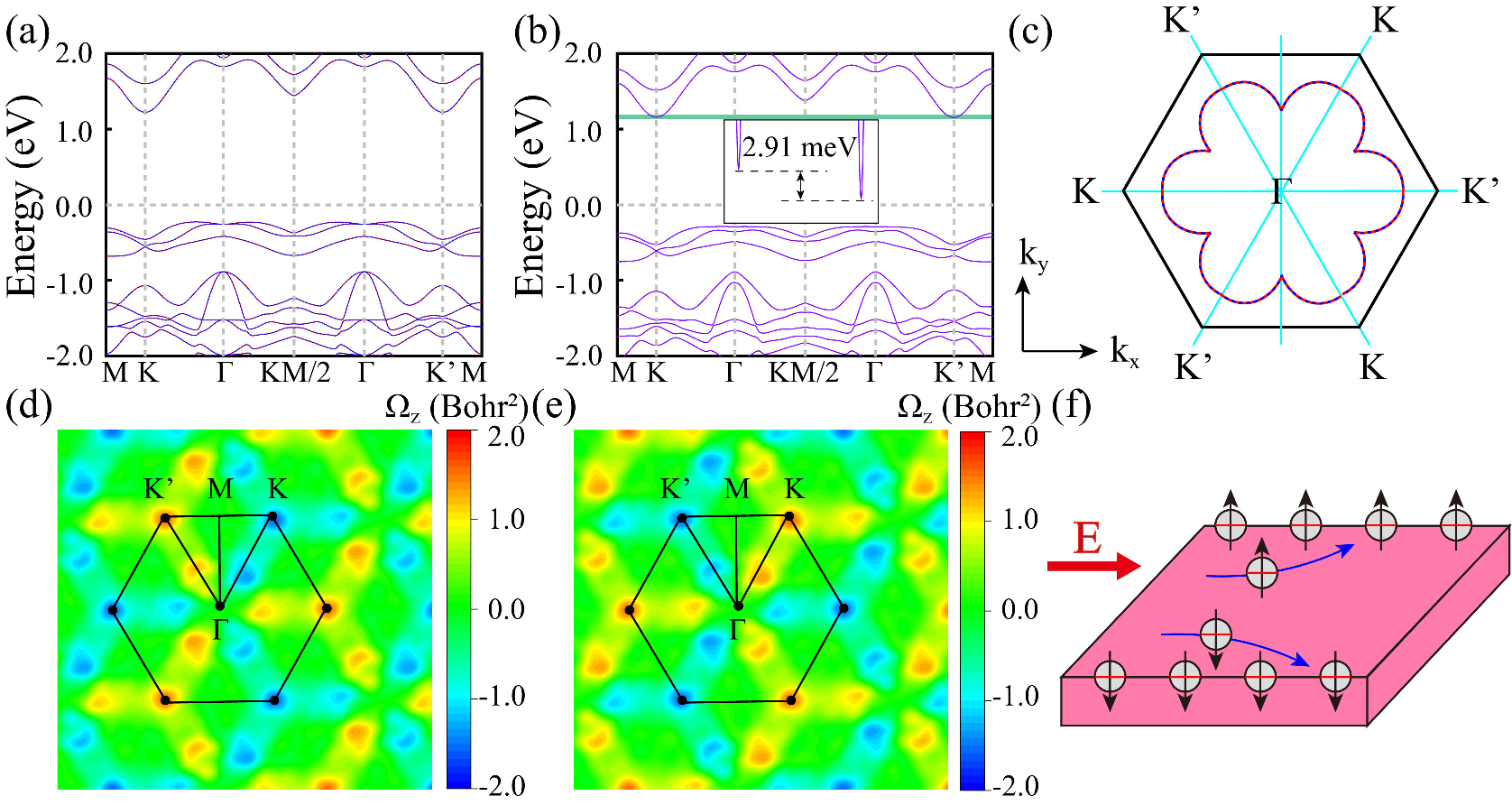}
\caption{(a) Spin-polarized band structure of monolayer VPSe$_3$. The solid red line and dashed blue line represent spin up and spin down bands, respectively. (b) The band structures with the SOC effect. The valley splitting of conduction band is exhibited by the green shading. (c) The valence band at around -1.2 eV. The Berry curvatures of monolayer VPSe$_3$ for (d) spin up state, and (e) spin down state in the entire Brillouin zone. (f) Schematic diagram of spin-valley Hall effect in the electron-doped monolayer VPSe$_3$ at the K' valley. The electrons and holes are shown by the - and + symbol. The $\uparrow$ and $\downarrow$ represents the spin up and spin down carriers, respectively.
}
\end{center}
\end{figure}

To understand the origin of valley splitting in monolayer VPSe$_3$, we employed $|$$\psi$$_c$$^{\tau}$$\rangle$=$\frac{1}{\sqrt{2}}$($|$p$_x$$\rangle$+i$\tau$$|$p$_y$$\rangle$)$\otimes$$|$$\uparrow$$\rangle$ as the orbital basis to build an effective Hamiltonian. The $\tau$ = $\pm$1 represent the valley index referring to the K/K' point. Taken the SOC effect as a perturbation term, the effective Hamiltonian can be written as
\begin{equation}
\hat{H}_{SOC} = \lambda \hat{S} \cdot \hat{L} = \hat{H}_{SOC}^{0} + \hat{H}_{SOC}^{1},
\end{equation}
where $\hat{S}$ and $\hat{L}$ are spin and orbital angular operators, respectively. The $\hat{H}_{SOC}^{0}$ and $\hat{H}_{SOC}^{1}$ denote the interaction between the same spin states and between opposite spin states, respectively. In monolayer VPSe$_3$, we only consider the interaction between spin up states. Thus, the $\hat{H}_{SOC}^{1}$ term can be neglected. Accordingly, the $\hat{H}_{SOC}^{0}$ can be rewritten by polar angles
\begin{equation}
\hat{H}_{SOC}^{0} = \lambda \hat{S}_{z'}(\hat{L}_zcos\theta + \frac{1}{2}\hat{L}_+e^{-i\phi}sin\theta + \frac{1}{2}\hat{L}_-e^{+i\phi}sin\theta),
\end{equation}
When the easy magnetization axis is along the out-of-plane, $\theta$ = $\phi$ = 0$^ \circ$, then the $\hat{H}_{SOC}^{0}$ term can be reduced as
\begin{equation}
\hat{H}_{SOC}^{0} = \lambda \hat{S}_{z} \hat{L}_z,
\end{equation}
Therefore, the energy level of the CBM valley can be written as E$_c$$^ \tau$ = $\langle$ $\psi$$_c$$^ \tau$ $|$ $\hat{H}$$_{SOC}^{0}$ $|$ $\psi$$_c$$^ \tau$ $\rangle$.
Consequently, the valley splitting can be described as
\begin{equation}
E_{c}^{K} - E_{c}^{K'} = i \langle p_x | \hat{H}_{SOC}^{0} | p_y \rangle - i \langle p_y | \hat{H}_{SOC}^{0} | p_x \rangle \approx \lambda,
\end{equation}
where the $\hat{L}_z|p_x \rangle$ = i$\hbar$$|p_y \rangle$, $\hat{L}_z|p_y \rangle$ = -i$\hbar$$|p_x \rangle$.

In addition, we understand valleys in terms of Berry curvature. Since monolayer AFM VPSe$_3$ is protected by $\emph{PT}$ symmetry, the Berry curvature is zero for the entire Brillouin zone. Therefore, we calculate the Berry curvature of spin up and spin down states. As shown in Fig. 2(d, e), the Berry curvature has equal magnitude at the K and K' points, while it shows opposite signs for the same valley of different spin states and different valleys of the same spin state. When the Fermi level shifts the conduction band of K' valley by electron doping, the spin up electrons from K' valley will be accumulated at the right edge of the sample, while the spin down electrons from K' valley will shift to the left edge under an in-plane electric field. As shown in Fig. 2(f), we name the phenomenon the spin-valley Hall effect.

\subsection{Electric field induce transformation into altermagnet}
From the perspective of device applications, the electric field stands out as the most effective means for tuning the physical quantity \cite{40,41}. This modulation method not only responds swiftly but also avoids causing damage to the materials themselves, demonstrating unique advantages in the application of micro- and nano-devices. Therefore, we investigate the effect of electric field on monolayer VPSe$_3$. Here, instead of re-optimizing VPSe$_3$ structure, we apply an electrostatic potential to monolayer VPSe$_3$. This causes the upper and lower Se atomic layer to be inequivalent, resulting in the broken $\emph{PT}$ symmetry. In the absence the SOC effect, the band structure holds sixfold rotational symmetry (C$_6$). The E(s, $\textbf{k}$) = E(-s, -$\textbf{k}$) relationship is satisfied on special high symmetric lines. However, the rotational symmetry of band structure will be reduced to threefold (C$_3$) under the SOC effect.

\begin{figure}[htb]
\begin{center}
\includegraphics[angle=0,width=1.0\linewidth]{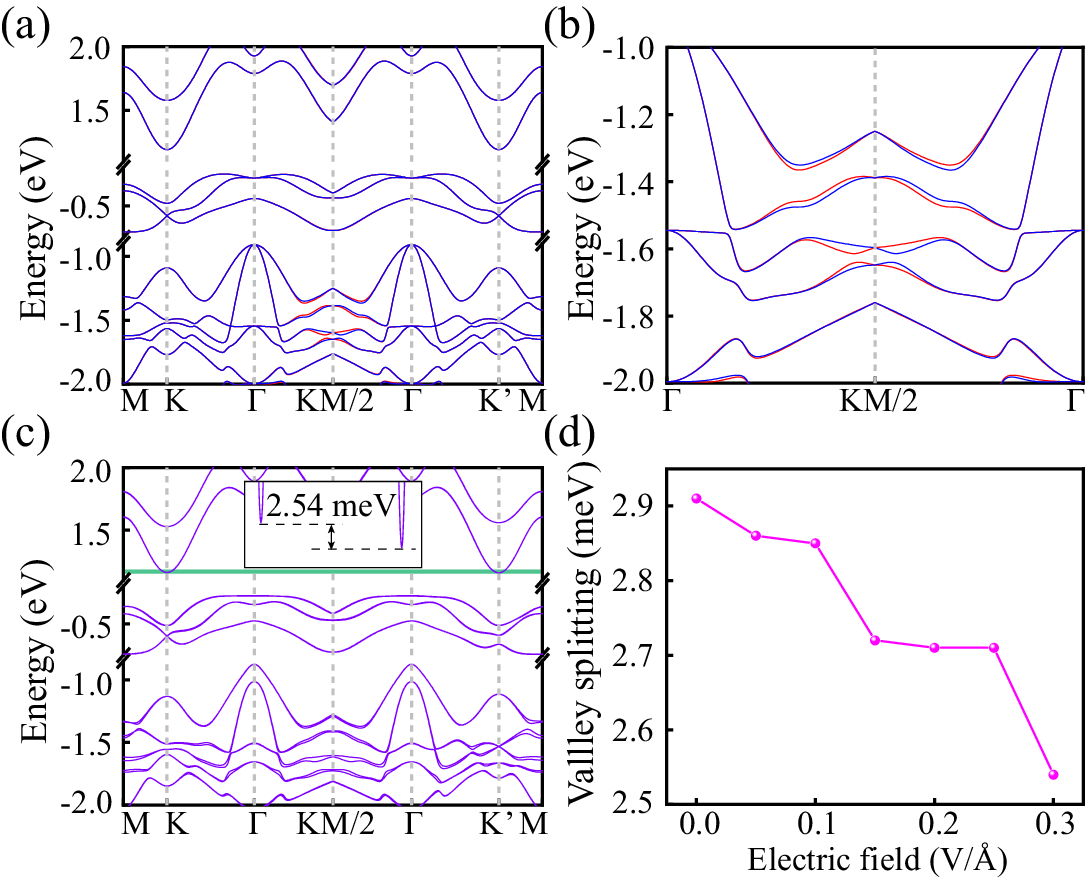}
\caption{(a) Spin-polarized band structure of monolayer VPSe$_3$ under an external electric field of 0.3 V/$\rm \AA$. The solid red and blue lines denote spin up and spin down bands, respectively. (b) An enlarged view of the region around KM/2 in (a). (c) The band structures with the SOC effect under an external electric field of 0.3 V/$\rm \AA$. The valley splitting of conduction band is shown by the green shading. (d) The valley splitting of monolayer VPSe$_3$ as a function of the external electric field.
}
\end{center}
\end{figure}

Firstly, we investigate the magnetic ground state at the 0.0 $\thicksim$ 0.3 V/$\rm \AA$. As shown in Fig. S2, the effect of electric field on the energy difference between FM and AFM states is only 0.1 meV, which means that the AFM phase is still the magnetic ground state. To confirm the symmetry analysis, we calculate the band structure at an electric field of 0.3 V/$\rm \AA$. As shown in Fig. 3(a), both valence and conduction bands of the valley at K and K' points is completely degenerate, which verifies that the band structure has C$_6$ symmetry without the SOC effect. More interestingly, as shown in Fig. 3(b), the spin splitting exhibits at the $\Gamma$-KM/2-$\Gamma$ path. It is the typical characteristic of altermagnet, which the AFM material doesn't has. Moreover, when the SOC is included, the valley splitting of 2.54 meV is still observed at the CBM. Since the valley degeneracy of K and K' points disappears, the symmetry decreases from the C$_6$ to C$_3$. It also demonstrates our symmetry analysis. Besides, as shown in Fig. 3(d), the change of valley splitting is only 0.4 meV at the 0.0 $\thicksim$ 0.3 V/$\rm \AA$. It is mainly because the splitting valley is contributed by the non-magnetic Se atom.

\subsection{Janus structure induce transformation into altermagnet}
The second approach is that it actually makes the upper and lower Se atomic layers unequal. We use S atom to replace lower Se atom to form the Janus structure V$_2$P$_2$S$_3$Se$_3$. To enrich the physical properties of 2D materials, Zhang $\emph{et al.}$ has successfully prepared Janus graphene as early as 2013 \cite{42}. After more than ten years of development, the current technology for preparing 2D Janus structures has been very mature and successfully synthesized Janus transition metal dichalcogenides (TMDs), such as MoSSe \cite{43}, WSSe \cite{44}, Janus MXenes \cite{45} etc. Due to the asymmetry between the lower and upper surfaces of the Janus structure, a significant built-in electric field is produced. Therefore, it naturally breaks the $\emph{PT}$ symmetry. We speculate that the same phenomenon of VPSe$_3$ applied electric field will be generated in V$_2$P$_2$S$_3$Se$_3$.

\begin{figure}[htb]
\begin{center}
\includegraphics[angle=0,width=1.0\linewidth]{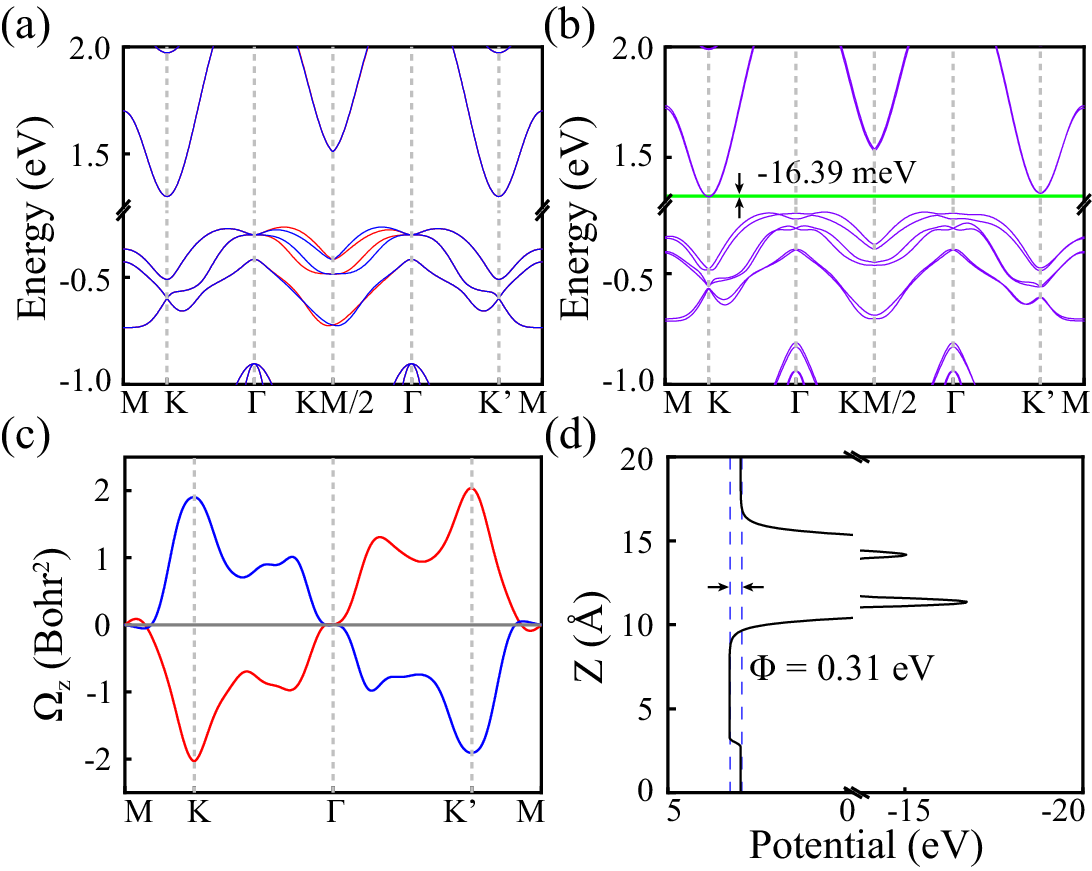}
\caption{(a) Spin-polarized band structure of Janus V$_2$P$_2$S$_3$Se$_3$. The solid red and blue lines represent spin up and spin down bands, respectively. (b) The band structures of Janus V$_2$P$_2$Se$_3$S$_3$ with the SOC effect. The valley splitting of conduction band is exhibited by the green shading. (c) The Berry Curvature along the high symmetry line. The red and blue lines represent Berry Curvatures of spin up and spin down states, respectively. (d) The planar averaged electrostatic potential of Janus V$_2$P$_2$S$_3$Se$_3$.
}
\end{center}
\end{figure}

In the absence of SOC, as shown in Fig. 4(a), the significant spin splitting is clearly observed in the $\Gamma$-KM/2-$\Gamma$ path. It indicates that the AFM to altermagnet transition is realized by constructing the Janus structure without the $\emph{PT}$ symmetry. In addition, the CBM at the K and K' points exhibits valley degeneracy. When the SOC is switched on, as illustrated in Fig. 4(b), the K and K 'valley degeneracies disappear, producing valley splitting up to -16.39 meV. It demonstrates that the intrinsic spontaneous valley polarization is achieved in the altermagnet monolayer V$_2$P$_2$S$_3$Se$_3$. Besides, to understand the valley related physics, we calculate Berry curvature of the monolayer V$_2$P$_2$S$_3$Se$_3$. As shown in Fig. 4(c), the Berry curvature of spin up exhibits a negative value near the K point and a positive near the K' point. On the contrary, the Berry curvature of spin down is completely reversed sign. It is also the characteristic of valley degeneracy. To further understand the valley splitting of monolayer V$_2$P$_2$S$_3$Se$_3$ -16.39 meV, we introduce the built-in electric field E$_{in}$ = ($\Phi_2$ - $\Phi_1$)/$\Delta h$, where $\Phi_1$, $\Phi_2$, and $\Delta h$ represent the electrostatic potential at the upper surface and lower surface, and structural height of V$_2$P$_2$S$_3$Se$_3$, respectively. As shown in Fig. 4(d), the built-in electric field is 0.10 V/$\rm \AA$ by the electrostatic potential. The built-in electric field of 0.10 V/$\rm \AA$ not only significantly increases the valley splitting, but also changes its sign. Why is the regulation effect of 0.10 V/$\rm \AA$ built-in electric field on valley splitting much better than that of external electric field. The important reason is the difference in SOC strength between S and Se atoms, which the S and Se atoms are 1.8 meV and 4.0 meV, respectively.

\subsection{Ferroelectric substrate induce transformation into altermagnet}
A third method to achieve altermagnets is by constructing heterostructures, which the heterostructure is naturally broken $\emph{PT}$ symmetry. In order to explore the regulation of substrate on valley splitting, we choose the ferroelectric substrate Al$_2$S$_3$ and VPSe$_3$ to form heterostructure. The lattice constant of Al$_2$S$_3$ is 3.59 $\rm \AA$. We find that the lattice mismatch rate of the 1 $\times$ 1 unit cell of VPSe$_3$ matching the $\sqrt{3}$ $\times$ $\sqrt{3}$ Al$_2$S$_3$ is only 0.3 $\%$. As shown in Fig. 5(a, e) and Fig. S3, four typical configurations are considered: top-S, top-Al, bridge, and hollow configuration. As listed in Table SI, the layer spacing in all configurations is 3.24 $\thicksim$ 3.82 $\rm \AA$, which indicates the interlayer is typically weak vdW interaction. In addition, all configurations exhibit an out-of-plane MAE, as listed in Table SII. From the relative total energy in Table SIII, the top-S, and top-Al are the most stable P$\uparrow$ and P$\downarrow$ configurations, respectively.

\begin{figure}[htb]
\begin{center}
\includegraphics[angle=0,width=1.0\linewidth]{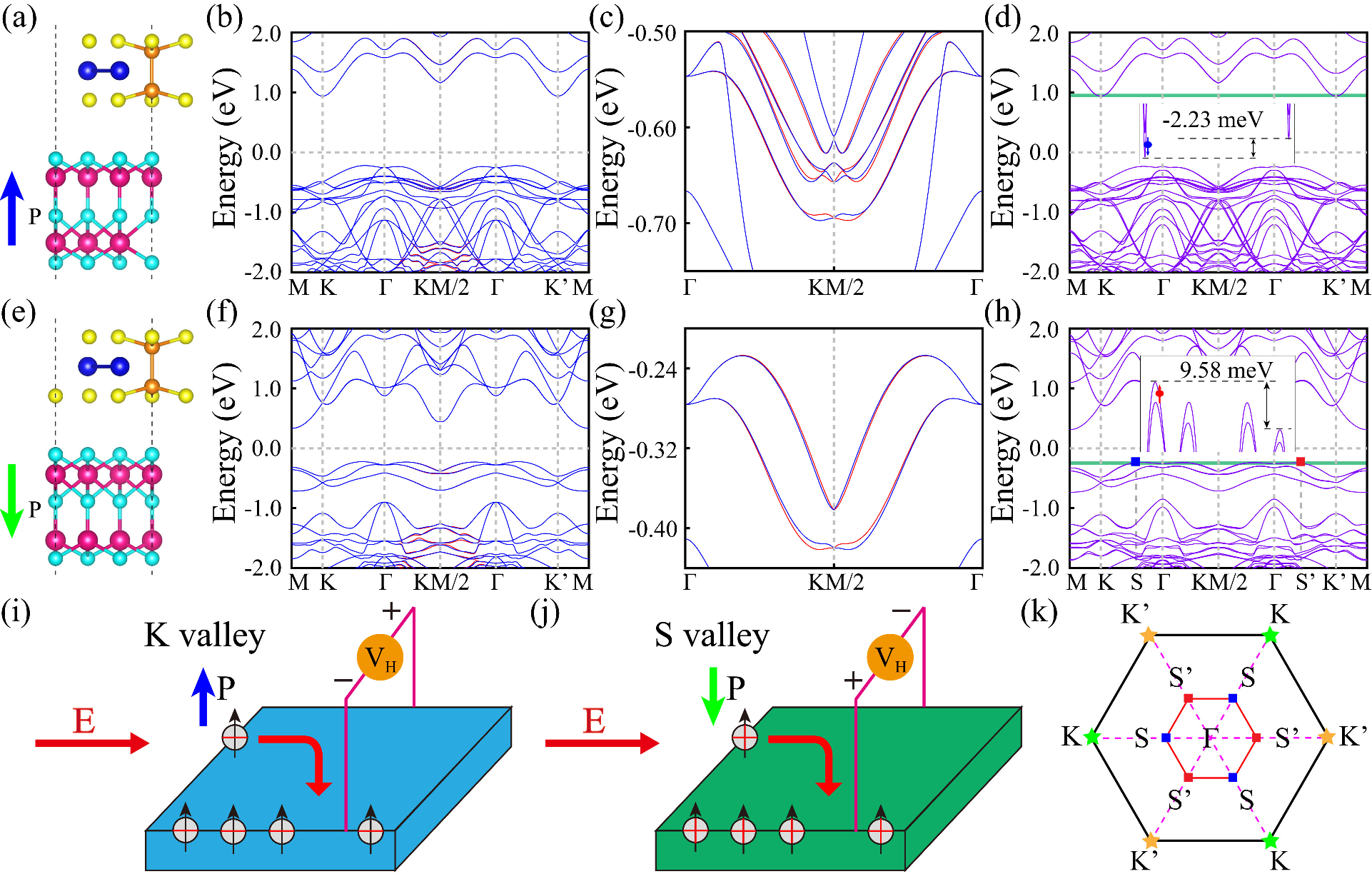}
\caption{ (a, e) Side views of VPSe$_3$/Al$_2$S$_3$ heterostructures with the top-S configuration. (a) and (e) correspond to the upward and downward polarization of Al$_2$S$_3$, respectively. (b, f) Spin-polarized band structure of VPSe$_3$/Al$_2$S$_3$ for (b) the P$\uparrow$ state and (f) the P$\downarrow$ state with the top-S configuration. The solid red and blue lines represent spin up and spin down bands, respectively. The (c) and (g) exhibit the enlarged view of the trhion around KM/2 for (b) and (f). (d, h) Band structure of VPSe$_3$/Al$_2$S$_3$ heterostructures with the SOC effect for (d) the P$\uparrow$ state and (h) the P$\downarrow$ state with the top-S configuration. The green shading show the valley splitting. (i, j) Schematic diagrams of anomalous valley Hall effect in (i) the electron-doped monolayer VPSe$_3$ at the K valley, and (j) the hole-doped at the S valley, respectively. The electrons and holes are represented by the - and + symbol. The $\uparrow$ and $\downarrow$ denotes the spin up and spin down carriers, respectively. (k) Location of K/K' and S/S' valleys in the first Brillouin zone.
}
\end{center}
\end{figure}

In the following, we take top-S configuration as an example for detailed analysis. When the ferroelectric polarization of Al$_2$S$_3$ is P$\uparrow$, the spin-polarized band structure is illustrated in Fig. 5(b). The VBM and CBM are located at K-$\Gamma$ path and K/K' point, simultaneously, the CBM of K and K' points exhibit valley degeneracy. When the P$\uparrow$ switches P$\downarrow$, as shown in Fig. 5(f), the CBM becomes the M point, while the VBM remains on the K-$\Gamma$ path. Very interestingly, the valley of VBM shows degeneracy, which is not a high symmetric point. Most importantly, as shown in Fig. 5(c, g), both P$\uparrow$ and P$\downarrow$ configurations produce spin splitting. This confirms the feasibility of our mechanism for realizing altermagnet. When the SOC is included, Fig. 5(d) shows the band structure of P$\uparrow$. The valley polarization direction is tuned by the Al$_2$S$_3$, however, the magnitude has barely changed. When the ferroelectric polarization of Al$_2$S$_3$ becomes P$\downarrow$, as shown in Fig. 5(h), the valley splitting of 9.58 meV occurs at the VBM.

To understand the origin of VBM and CBM valley splitting, as shown in Fig. S6, we calculate the orbital-resolved band structure of top-S configuration VPSe$_3$/Al$_2$S$_3$. The CBM valley splitting comes from the Se atom liked monolayer VPSe$_3$, while the VBM valley splitting is contributed by d$_{x^2-y^2}$+d$_{xy}$ of V atoms. We uses $|$$\psi$$_v$$^{\tau}$$\rangle$=$\frac{1}{\sqrt{2}}$($|$d$_{xy}$$\rangle$+i$\tau$$|$d$_{x2-y2}$$\rangle$)$\otimes$$|$$\uparrow$$\rangle$ as the orbital basis to construct an effective Hamiltonian. The derivation is consistent. Here, we only give the conclusion.
\begin{equation}
E_{v}^{S} - E_{v}^{S'} = i \langle d_{xy} | \hat{H}_{SOC}^{0} | d_{x2-y2} \rangle - i \langle d_{x2-y2} | \hat{H}_{SOC}^{0} | d_{xy} \rangle \approx 4\alpha,
\end{equation}
where the $\hat{L}_z|d_{xy} \rangle$ = -2i$\hbar$$|d_{x2-y2} \rangle$, $\hat{L}_z|d_{x2-y2} \rangle$ = 2i$\hbar$$|d_{xy} \rangle$, and $\alpha = \lambda \langle d_{x2-y2} |\hat{S}_{z'}| d_{x2-y2} \rangle$.

It should be pointed out that the valley-dependent spin splitting occurs with the lower symmetry and SOC. Therefore, the anomalous valley Hall effect will be observed. As shown in Fig. 5(i), the spin down electron of K valley will be generated and accumulate on one boundary in electron doping condition for the P$\uparrow$. On the contrary, in the hole doping case, the spin up hole of S valley will also be produced and accumulate on same boundary for the P$\downarrow$ [see Fig. 5(j)]. Furthermore, Fig. 5(k) shows the location of valley splits under two polarization states in the first Brillouin zone. All configurations of valley splitting are listed in Table I. Noted that the same phenomenon appears for other configurations, as shown in Fig. S4 and Fig. S5. It means that ferroelectric substrate Al$_2$S$_3$ can effectively tune the direction and magnitude of VPSe$_3$ valley polarization. More interestingly, the ferroelectric polarization can switch altermagnetism effect, which indicates the ferroelectric polarization is coupled to the altermagnetic spin splitting.

\begin{table}[htbp]
\caption{
Calculate the valley splitting of different VPSe$_3$/Al$_2$S$_3$ heterostructure configurations, which is defined as valley splitting = E$_{\rm K}$ - E$_{\rm K'}$. }
\begin{tabular}{cccccccc}
	\hline
	                                           &  top-S       & top-Al      & hollow     & bridge   \\
	\hline
	 VPSe$_3$/Al$_2$S$_3$ P$\uparrow$ (meV)    &  -2.23       & -3.45       & -3.13      & -2.82    \\
    \hline
     VPSe$_3$/Al$_2$S$_3$ P$\downarrow$ (meV)  &   9.58       &  9.15       &  8.74      &  8.96    \\
    \hline
\end{tabular}
\end{table}	

\section{CONCLUSION}
In summary, we present a novel strategy to realizing altermagnet in 2D material. We use an electric field, Janus structure, and ferroelectric substrate to make the upper and lower atoms asymmetric, while the $\emph{PT}$ symmetry of the AFM state is broken, resulting in a transition to altermagnetic state. Based on the symmetry analysis and DFT calculation, the mechanism is proved to be feasible in monolayer VPSe$_3$. The magnetic ground state of monolayer VPSe$_3$ is an AFM state, which exhibits a spontaneous valley splitting of 2.91 meV. The electric field, the Janus structure, and the ferroelectric substrate can effectively break the $\emph{PT}$ symmetry, so that the band structure can exhibit spin splitting at the $\Gamma$-KM/2-$\Gamma$ path. It proves that multiple-physical-fields can transform monolayer VPSe$_3$ from AFM to altermagnetic state. Moreover, the electric field can adjust the magnitude of valley splitting, while the Janus structure can regulate both the magnitude and direction of valley splitting. For the ferroelectric substrate Al$_2$S$_3$, the valley polarization direction is effectively tuned and the magnitude hardly changes under the P$\uparrow$. However, when the ferroelectric polarization of Al$_2$S$_3$ switches from P$\uparrow$ to P$\downarrow$, the valley splitting significantly increases. In addition, the ferroelectric polarization can switch altermagnetic effect and realize anomalous valley Hall effect. Our work not only provides a route to realize the altermagnet in 2D system, but also an efficient ways to tune valley splitting.

\section*{ACKNOWLEDGEMENTS}
This work is supported by the National Natural Science Foundation of China (Grants No. 12474238, and No. 12004295). P. Li also acknowledge supports from the China's Postdoctoral Science Foundation funded project (Grant No. 2022M722547), the Fundamental Research Funds for the Central Universities (xxj03202205), and the Open Project of State Key Laboratory of Silicon and Advanced Semiconductor Materials (No. SKL2024-10), and the Open Project of State Key Laboratory of Surface Physics (No. KF2024$\_$02).


\end{document}